\documentclass[12pt]{article}
\usepackage{enumerate,amssymb}
\renewcommand{\a}{\alpha}
\renewcommand{\b}{\beta}
\newcommand{\g}{\gamma}           
\renewcommand{\d}{\delta}

\newcommand{\ka}{\kappa}

\newcommand{\la}{\lambda}

\newcommand{\s}{\sigma}           
         
\newcommand{\f}{{\phi}}

\newcommand{\eps}{{\epsilon}}

\newcommand{\be}{\begin{equation}}
\newcommand{\ee}{\end{equation}}
\newcommand{\eqn}[1]{\label{#1}\end{equation}}

\newcommand{\bea}{\begin{eqnarray}}
\newcommand{\eea}{\end{eqnarray}}
\newcommand{\eqan}[1]{\label{#1}\end{eqnarray}}

\newcommand{\ba}{\begin{array}}
\newcommand{\ea}{\end{array}}

\newcommand{\nn}{\nonumber}

\begin{document}

\begin{center}
{\bf   Non-Minimal String Corrections And Supergravity}\\[14mm]

S. Bellucci\\

{\it INFN-Laboratori Nazionali di Frascati\\
Via E. Fermi 40, 00044 Frascati, Italy\\ mailto: bellucci@lnf.infn.it}\\[6mm]

D. O'Reilly\\

{\it Physics Department, The Graduate School and University
Center\\365 Fifth Avenue
New York, NY 10016-4309\\ mailto: doreilly@gc.cuny.edu}\\[6mm]

\end{center}

\vbox{\vspace{3mm}}

\begin{abstract}

We reconsider the well-known issue of string corrections to
Supergravity theory. Our treatment is carried out to second order in
the string slope parameter. We establish a procedure for solving the
Bianchi identities in the non minimal case, and we solve a long
standing problem in the perturbative expansion of D=10, N=1 string
corrected Supergravity, obtaining the H sector tensors, torsions and
curvatures. \vbox{\vspace{1mm}}

\end{abstract}

\vbox{\vspace{3cm}}


\newpage

%%%%%%%%%%%%%%%%%%%%%%%%%%%%%%%%%%%%%%%%%%%%%%%%%%%%%%%%%%%
\section{Introduction}

In view of the renewed interest in string corrected D=10, N=1
Supergravity \cite{1}, we revisit an outstanding problem concerning
the case to second order in the perturbative expansion. String
corrected D=10 and N =1 supergravity is believed to be the low
energy limit of string theory, \cite{1}-\cite{4}. Some years ago a
program was developed by Gates and collaborators to incorporate
string corrections into the supergravity equations of motion
\cite{2}-\cite{4}. This approach solved the problem of maintaining
manifest supersymmetry. Recently the bosonic equations of motion for
D=10, N=1 supergravity fields at superspace and component levels
have been obtained and have been shown to be derivable from a
lagrangian \cite{1}. The authors have done this to first order in
the string slope parameter perturbatively. A long standing problem
has been that of obtaining the successful closure of the Bianchi
identities to second order. It was stated in \cite{5} that
proceeding to higher than first order would yield interesting
results. However in practice doing so was not so
straightforward.\footnote{The suggestion of changing the torsion
constraints was given in \cite{51}. At the end of this paper a
warning appeared: accepting that everything works well and the
solution proposed by the authors is consistent, it might be that it
differs from the one in ref. \cite{5} simply by a field
redefinition.}

In the present work we establish a procedure for solving the Bianchi
identities to second order and we solve the long standing problem of
closure in the H sector. It was suggested that a second order
solution would require a modification of the torsion
$T_{\a\b}{}^{g}$ to second order, \cite{1}, containing the so called
X tensor. We propose an Ansatz for the X tensor and show that it
allows for closure in the H and Torsion sectors. We also find the
relevant torsions and show mutual consistency. In the curvature
sector we identify $R^{(2)}{}_{\a\b de}$. Crucial to this solution
are the results outlined in section (4).

The perturbative approach by Gates and coworkers is well documented and
discussed in the literature, and we will not recount it here. For a
recent review and for an up to date commentary see \cite{1}, and
references therein. Our starting point will be the Bianchi
identities as listed in \cite{5}. The sigma matrix identities and
symmetries are recorded in \cite{3}.

These geometrical methods nowadays are known as deformations
\cite{1}, and the constraints have sometimes been referred to in the
past as beta function favored ($\b FF$) constraints \cite{vash} (see
for related subjects e.g. \cite{conf2}). In the past, such methods
allowed for the determination of the most general higher derivative
Yang-Mills action to order $\gamma^3$, which is globally
supersymmetric and Lorentz covariant in D=10 spacetime (see e.g.
\cite{more}), a result which is important for topologically
nontrivial gauge configurations of the vector field, e.g., for
compactifield string theories on manifolds with topologically
nontrivial properties.

This paper is organized as follows. In section (2) we set up the
Bianchi identities. In section (3) we discuss some implications of
the X tensor cited in \cite{1}. In section (4) we outline key
results necessary for obtaining the solutions. As the derivations
are lengthy we have not included them in this letter. In section (5)
we give the H sector solution through identifying
$T^{(2)}{}_{\a\b}{}^{\g}$ and by the elimination of non linear
terms. In section (6) we propose a candidate for the so called X
tensor and show that it achieves closure in the H sector by a
different method and that it also admits solutions to the torsion
equations. The remaining sections examine further checks and
curvatures, finally concluding.

\section{The Bianchi identities}
The full set of Q and G Bianchi identities is listed in ref.
\cite{5}. Below we list only those which we need to consider here.
The $H$ tensor is related to the $G$ tensor in superspace as
follows:

 \bea
G_{ADG}~= H_{ADG}~+~\g Q_{ADG}+ \b Y_{ADG} \eea

\noindent where Q is the Lorentz-Chern Simons superform, $Y_{ADG}$
is the Yang Mills Superform, and $\g$ is proportional to the string
slope parameter. That is, the action for massless fields of
heterotic or type I superstrings may be expanded, with $\b$ set to
zero, as follows \cite{2}:
 \bea S_{eff}~=~\frac{1}{\ka^{2}}\int
d^{10}x
e^{(-1)}[\emph{L}_{(0)}~+~\sum_{n=1}^{n=\infty}(\gamma)^{n}\emph{L}_{(n)}]
\eea
 Within the framework of the Bianchi
identities we have a perturbative prescription that will allow us to
incorporate string corrections into the theory, and maintain it
manifestly supersymmetric. We first solve the identities satisfied
by the H tensor. The first three are
 \bea
\frac{1}{6}\nabla_{(\a|}H_{|\b\g\d)}~-~\frac{1}{4}T_{(\a\b|}{}^{E}
H_{E|\g\d)}~=~(-\frac{\g}{4})R_{(\a\b|ef}R_{|\g\d)}{}^{ef} \eea
 \bea
{\frac{1}{2}}\nabla_{(\a|}H_{|\b\g)d}~-~{\nabla_{d}}H_{\a\b\g}~
-~\frac{1}{2}T_{(\a\b|}{}^{E}H_{E|\g)d}~
+~\frac{1}{2}T_{d(\a|}{}^{E}H_{E|\b\g)}~=~(-\g)R_{(\a\b|ef}R_{|\g)d}{}^{ef}
\eea

 \bea~ \nabla_{(\a|}H_{\b)
cd}~+~\nabla_{[c|}H_{|d]\a\b}~-~T_{\a\b}{}^{E}H_{Ecd}~
-~T_{cd}{}^{E}Q_{E\a\b}~-~T_{(\a|[c|}{}^{E}H_{E|d]|\b)}~=~\nn \\
~-\g[2R_{\a\b ef}R_{cd}{}^{ef}~
+~R_{(\a|[c|}{}^{ef}R_{|d]|\b)}{}_{ef}]\eea

We also require the torsions

\bea
T_{(\a\b|}{}^{\la}T_{|\gamma)\la}{}^{d}~-~T_{(\a\b|}{}^{g}T_{|\gamma)g}{}^{d}~
-~\nabla_{(\a|}T_{\b\g)}{}^{d}~=~0 \eea and

 \bea
T_{(\a\b|}{}^{\la}T_{|\g)\la}{}^{\d}~-~T_{(\a\b|}{}^{g}T_{|\g)g}{}^{\d}
~-~\nabla_{(\a|}T_{|\b\g)}{}^{\d} -~\frac{1}{4}R_{(\a\b|
de}\s^{de}{}_{|\g)}{}^{\d} ~=~0 \eea

The first order solution, originally given in \cite{5}, was recently
recalculated \cite{1}. Many avenues such as null spaces were tried
in order to proceed to second order, but without success. It was
suggested that a generalization of the torsion $T_{\a\b}{}^{g}$
would be necessary in order to proceed to second order \cite{1}. One
of the problems at hand therefore is to find the form of this
generalization, known as the X tensor. Here we do this and we also
find the second order torsions $T^{(2)}{}_{\a\b}{}^{g}$,
$T^{(2)}{}_{\a\b}{}^{\la}$, $T^{(2)}{}_{\a b}{}^{g}$, $T^{(2)}{}_{\a
b}{}^{\g}$ and curvatures $R^{(2)}{}_{\a\b de}$ and $R^{(2)}{}_{\a
bde}$

We check our results by showing mutual consistency.

At this stage we will draw attention to our simple second order
notation. The superscript in brackets refers to the second order
quantities. For brevity we will not relist the first order
quantities as found in \cite{1} and \cite{5} unless necessary. Hence
we have, for example

\bea R_{\a\b de} = R^{(0)}{}_{\a\b de} + R^{(1)}{}_{\a\b de}+
R^{(2)}{}_{\a\b de} +....\eea

\noindent where $R^{(0)}{}_{\a\b de}$ and $R^{(1)}{}_{\a\b de}$ are
listed in \cite{1} and \cite{5}. To begin with, $H_{\b\g\d}$ is set
to zero as in \cite{1} and \cite{5}. We have not seen that it is
required to be other than zero to close the Bianchi identities. We
have seen that if it is non zero the H sector Bianchi identities
fail to close.

\section{The Problem}
Imposing the conventional constraint
$T^{g}{}_{\a\b}~=~i\s^{g}{}_{\a\b}$ to all orders
leads to failure to close the Bianchi identities to second order. In
the present case that constraint is modified. From the conventional
constraints listed in \cite{1} the most general form of the zero
dimensional torsion is
 \bea T_{\a\b}{}^{g}~=~i\s^{g}{}_{\a\b}
~+~\s^{pqref}{}_{\a\b}X_{pqref}{}^{g} \eea
 Here we absorb the coefficient $\frac{i}{5!}$ used in ref. \cite{1} into the X tensor.

 Earlier, because of the existence of an apparently intractable
term which arose in the H sector Bianchi identities, closure could
not be obtained with $T_{\a\b}{}^{g}=i\s^{g}{}_{a\b}$. At the time
an unsolvable term could not be incorporated into the torsion
$T^{(2)}{}_{\a\b}{}^{\la}$, which would have allowed for a solution.
In the following we see that the X tensor must contribute to second
order.
 Let us consider the Bianchi identity at dimension one-half.
 If the X tensor is zero, then we have,
using the constraints in \cite{5},  \bea
T^{(2)}{}_{(\a\b|}{}^{\la}\s_{|\gamma)\la}{}^{d}~=~\s_{(\a\b|}{}^{g}T^{(2)}{}_{|\gamma)g}{}^{d}
 \eea
 Therefore $T^{(2)}{}_{\a\b}{}^{\la}$ must have a similar sigma matrix structure to the RHS of (10).
 We know from the H sector Bianchi identities that
 $H^{(2)}{}_{g \g d}$ satisfies an equation of the form \cite{6}
\bea
\s^{g}{}_{(\a\b|}H^{(2)}{}_{g|\g)d}~~~=~~~~~~~~~~~~~~~~~~~~~~~~~~~~~~~~~~~~~~~~~~~~~~~~~\nn\\
\s^{g}_{(\a\b|}[M_{g|\g)d}]+~\s^{g}{}_{(\a\b|}T^{(2)}{}_{|\g) gd}
-~~\s_{d(\a|\la}T^{(2)}{}_{|\b\g)}{}^{\la}~
-~8\g^{2}{\s_{e(\a|\eps}}{\s_{f(\b|\tau}}{\s_{d|\g)\f}}
T^{(0)}{}_{kp}{}^{\eps}T^{(0)}{}^{kp}{}^{\tau}T^{(0)}{}^{ef}{}^{\f}\nn\\
~~~~~~~~~~~~~~~~~~~~~~~~~~~~~\eea Looking at (11) we see that we
will encounter an intractable term in the H sector unless we can
absorb it into the $T^{(2)}{}_{\b\g}{}^{\la}$ term. There is no
known sigma matrix identity that will allow the last term on the RHS
of (11) to be written in the form necessary to solve for
$H^{(2)}{}_{g \g d}$. Let us call it the $T^{3}$ term. We see that
the $T^{3}$ term has the same sigma matrix structure as the
$T^{(2)}{}_{\b\g}{}^{\la}$ term. We consider the option of equating
the $T^{(2)}{}_{\b\g}{}^{\la}$ term with the $T^{3}$ term. This is
possible only when X is non zero, and it will constitute one of the
many necessary components of the solution.

\section{Results Necessary for the Solution}
The following combination of results led to the solution, in
combination with other observations. The first requirement will be
the X tensor Ansatz. We will require the zeroth order
spinor derivative of $T^{(0)}{}_{ef}{}^{\d}$, available from
\cite{5},

\bea
\nabla_{\g}T^{(0)}{}_{ef}{}^{\d}=-\frac{1}{4}\s^{mn}{}_{\g}{}^{\d}R_{efmn}
~+~T^{(0)}{}_{ef}{}^{\la}T^{(0)}{}_{\g\la}{}^{\d} \nn\\\eea

 We refer to its $\chi$ part and its curvature part using
 a suitable subscript as in eqs. (13)and (15) below. We have the following important results,
  the derivations of which are lengthy and not included in
this letter \cite{6}

\bea
~T^{(0)}{}_{(\a\b|}{}^{\la}\frac{i\g}{6}\s^{pqref}{}_{|\g)\la}A^{(1)}{}_{pqr}H^{(0)}{}_{def}
-\frac{i\g}{6}\s^{pqref}{}_{(\a\b|}[H^{(0)}{}_{def}\nabla_{|\g)}A^{(1)}{}_{pqr}]\mid_{(\chi)}\nn\\
=~+~i\g\s^{g}_{(\a\b|}H^{(0)}{}_{def}\nabla_{|\g)}A^{(1)}{}_{gef}\mid_{(\chi)}
\eea

We also require the following:

 \bea
+~\frac{i\g}{6}\s^{pqref}{}_{(\a\b|}H^{(0)}{}_{def}\nabla_{|\g)}A^{(1)}{}_{pqr}|_{(R)}
=~ \frac{\g^{2}}{12}\s^{pqref}{}_{(\a\b|}H^{(0)}{}_{def}
\s_{pqr}{}_{\eps\tau}T^{(0)}{}_{kp}{}^{\eps}\s^{mn}{}_{|\g)}{}^{\tau}R^{(0)}{}_{mn}{}^{kp}\nn\\
=~\frac{\g^{2}}{2}\s^{g}{}_{(\a\b|}\s_{gef}{}_{\eps\tau}T^{(0)}{}_{kp}{}^{\eps}\s^{mn}{}_{|\g)}{}^{\tau}
R^{(0)}{}_{mn}{}^{kp}H^{(0)}{}_{d}{}^{ef}
+~16\g^{2}\s^{g}{}_{(\a\b|}\s_{e|\g)\eps}T^{(0)}{}_{kp}{}^{\eps}R^{(0)}{}_{fg}{}^{kp}H^{(0)}{}_{d}{}^{ef}\nn\\
~\eea For later transparency, using the identity
 $\s^{g}{}_{(\a\b|}\s_{g |\g)\la}=0$, this can be rewritten as
\bea
~\frac{i\g}{6}\s^{pqref}{}_{(\a\b|}H^{(0)}{}_{def}\nabla_{|\g)}A^{(1)}{}_{pqr}|_{(R)}\nn\\
=-i\g\s^{g}{}_{(\a\b|}[\nabla_{|\g)}A^{(1)}{}_{gef}]|_{(R)}H^{(0)}{}_{d}{}^{ef}+
4i\g \s^{g}{}_{(\a\b|}R^{(1)}{}_{|\g) gef}H^{(0)}{}_{d}{}^{ef}\eea

Combining (13) and (15) we get a crucial equation which will
facilitate finding our solution.  In a very conveniently reduced
form, what once seemed apparently intractable, will become

\bea
~+\frac{i\g}{6}T^{(0)}{}_{(\a\b|}{}^{\la}\s^{pqref}{}_{|\g)\la}A^{(1)}{}_{pqr}H^{(0)}{}_{def}-
\frac{i\g}{6}\s^{pqref}{}_{(\a\b|}H^{(0)}{}_{def}\nabla_{|\g)}A^{(1)}{}_{pqr}\nn\\
=+i\g\s^{g}{}_{(\a\b|}[\nabla_{|\g)}A^{(1)}{}_{gef}]H^{(0)}{}_{d}{}^{ef}-
4i\g\s^{g}{}_{(\a\b|}R^{(1)}{}_{|\g) gef}H^{(0)}{}_{d}{}^{ef}\eea

Along with the sigma matrix identities listed in \cite{3}, we also
require the important result that
 \bea
\s^{pqref}{}_{(\a\b|}\s_{e|\g)\f}=~=~-~\s^{pqref}{}_{\f(\a|}\s_{e|\b\g)}
\eea The latter allows us to write

\bea \frac{i\g}{12}\s^{pqref}{}_{(\a\b|}A^{(1)}{}_{pqr}
R^{(0)}{}_{|\g)}{}_{def}\nn\\=~+\frac{\g}{6}\s^{g}{}_{(\a\b|}\s^{pqre}{}_{g}{}_{|\g)\f}A^{(1)}{}_{pqr}T^{(0)}{}_{d
e}{}^{\f} \eea

Closure was not immediately evident in the curvature sector. To
achieve this equations (58) and (72) were also required \cite{6}.
\section{The H Sector Solution}

It is a straightforward to solve for $H^{(2)}{}_{g\g\d}$. To do so
we use the constraints in ref. \cite{5} and substitute them into
equation (3). We then extract a sigma matrix coefficient from each
term and solve. After some algebra we obtain

 \be H^{(2)}{}_{g \g\d }~~=~[{-4\g}H^{(0)}{}_{gef}R^{(1)}{}_{\g\d}{}^{ef}~
~-~~\frac{1}{2}T^{(2)}{}_{g\g\d}] \ee \ We can also write the latter
expression as follows:
 \bea
H^{(2)}{}_{\a\b d}~
=~\s_{\a\b}{}^{g}[{8i\g}H^{(0)}{}_{def}L^{(1)}{}_{g}{}^{ef}~ -~i\g
H^{(0)}{}_{def}A^{(1)}{}_{g}{}^{ef}]\nn\\~+~
\s^{pqref}{}_{\a\b}[\frac{-i\g}{6}H^{(0)}{}_{def}A^{(1)}{}_{pqr}~-
~\frac{1}{2}X_{pqrefd}] \eea

For the next Bianchi identity, applying the constraints of ref.
\cite{5} to (4) reduces it to

\bea \frac{+i}{2}\s^{g}{}_{(\a\b|}H^{(2)}{}_{g|\g)d}~=~
{\frac{1}{2}}\nabla_{(\a|}H_{|\b\g)d}{}^{Order\g^{2}}-~
\frac{1}{2}T^{(0)}{}_{(\a\b|}{}^{\la}H^{(2)}{}_{\la|\g)d}~\nn\\
-~i2\g\s^{g}{}_{(\a\b|}[\Pi^{(1)}{}_{g}{}^{ef}R^{(0)}{}_{|\g)def}+
H^{(0)}{}_{gef}R^{(1)}{}_{|\g)d}{}^{ef}]\nn\\
+~\frac{i\g}{24}\s^{pqr}{}_{ef}{}_{(\a\b|}R^{(0)}{}_{|\g)d}{}^{ef}A^{(1)}{}_{pqr}
~-~\frac{i}{4}\s_{d(\a|\la}T^{(2)}{}_{|\a\b)}{}^{\la}+~\frac{i}{4}\s^{g}_{(\a\b|}T^{(2)}{}_{d|\g)g}
\nn\\\eea

 where
 \bea
 \Pi^{(1)}{}_{g}{}^{ef}=L^{(1)}{}_{g}{}^{ef}
-~\frac{1}{8}A^{(1)}{}_{g}{}^{ef} \eea We write it in the above form
for transparency and for future reference. In particular we note
that the last two terms in (21) can be obtained using equation (6),
and that they can be written in terms of the X tensor. Equation (21)
contains a proliferation of non solvable $\chi$ terms. The X tensor
is also contained within this equation. It is lengthy and does not
advance our argument to write it out in full here.

We will solve this Bianchi identity by two routes. Firstly
we solve it by eliminating the X tensor, and by identifying
$T^{(2)}{}_{\a\b}{}^{\la}$. We need to consider the spinor
derivative of $H_{\b\g d}{}^{Order\g^{2}}$. We do not yet know the
form of the X tensor, so we employ a torsion and a curvature to
eliminate $\chi$ and X tensor terms. We will later calculate it
directly and show agreement. We must remember to include the second
order derivative contributions which come from the first order
result. To first order, \cite{5}, we have
 \bea H_{\b \g
d}~=~\frac{i}{2}\s_{d\b\g}+i4\g\s^{g}{}_{\b\g}H^{(0)}{}_{gef}H^{(0)}{}_{d}{}^{ef}
\eea As pointed out in \cite{vash}, the field independence of the
leading term in the $H_{\alpha\beta c}$ component of the 3-form
field strength is indicative of the correspondence of these
equations in the limit $\gamma =0$, to the superspace geometry in a
string-frame description of pure supergravity. Taking the spinor
derivative of this first order term will generate second order
contributions. We have from first order, which we leave as is, \bea
\frac{1}{2}\nabla_{(\a|}H^{(1)}{}_{|\b\g) d}{}^{(Order\g^{2})}~~=~~
\s^{g}_{(\a\b|}({2i\g}\nabla_{|\g)}[H^{(0)}{}_{gef}H^{(0)}{}_{d}{}^{ef}]^{Order(1)})
\eea
 We seek to solve for $H^{(2)}{}_{g\g d}$ in an
expression of the form

 \bea
\s^{g}{}_{(\a\b|}H^{(2)}{}_{g|\g)d}~=~\s^{g}{}_{(\a\b|}M^{(2)}{}_{g|\g)d}
\eea
 Finally we extract the correct expression for $H^{(2)}{}_{g\g d}$
with an appropriate symmetrization operator, (77).

We need to take the derivative of equation (19). To do this we use a
first order curvature and the dimension one-half torsion to second
order. This will eliminate the $\chi$ terms in (21), and it will
also isolate the torsion $T^{(2)}{} _{\a\b}{}^{\la}$.

The first order curvature required is \bea
\nabla_{(\a|}R^{(1)}{}_{|\b\g)ef}~~=~~~
T^{(0)}{}_{(\a\b|}{}{}^{~\la}R^{(1)}{}_{|\g)\la ef}
~+~T^{(1)}{}_{(\a\b|}{}{}^{~\la}R^{(0)}{}_{|\g)\la ef}
~-~T^{(0)}{}_{(\a\b|}{}^{~g}R^{(1)}{}_{|\g)gef}~\nn\\
-~T^{(1)}{}_{(\a\b|}{}{}^{~g}R^{(0)}{}_{|\g)gef}~ \eea

The required torsion is given in equation (6). This gives the
complete expression for the derivative
 \bea \frac{1}{2}\nabla_{(\a|}H_{|\b\g)d}{}^{Order 2}~=~
\s^{g}_{(\a\b|}[2i\g\nabla_{|\g)}(H^{(0)}{}_{def}H^{(0)}{}_{g}{}^{ef})^{Order
2}~+~4i\g [\nabla_{|\g)}H^{(0)}{}_{def}]\Pi^{(1)}{}_{g}{}^{ef}]~~~~~~~~~~~~~~~~~\nn\\
~-~\frac{i\g}{12}
\s^{pqref}{}_{(\a\b|}[\nabla_{|\g)}H^{(0)}{}_{def})]A^{(1)}{}_{pqr}~~
+~\frac{1}{2}\chi_{(\a|}T^{(2)}{}_{|\b\g)}{}_{d}~
-~\frac{i}{4}\s_{d(\a|\la}T^{(2)}{}_{|\b\g)}{}^{\la}~~~~~~~~~~~~\nn\\
+~\s^{g}_{(\a\b|} [\frac{1}{4}
\s_{g}{}^{\la\f}\chi_{\f}T^{(2)}{}_{|\g)\la d}
~+~\frac{i}{4}T^{(2)}{}_{|\g)gd}]~~~~~~~~~~~~~~~~~\nn\\
+~\s^{g}{}_{(\a\b|}[2\g\s_{g}{}^{\la\f}\chi_{\f}
R^{(1)}{}_{|\g)\la}{}_{ef} H^{(0)}{}_{d}{}^{ef}~+~2i\g
H^{(0)}{}_{def}R^{(1)}{}_{|\g)g}{}^{ef}] \nn\\+~4\g
H^{(0)}{}_{def}\chi_{(\a|}R^{(1)}{}_{|\b\g)}{}^{ef}~~~~~~~~~~~~~~~~~\nn\\
\eea

After substitution of the derivative term, (27), into (21) we
eliminate the non linear terms and after many cancelations we arrive
at

\bea \frac{i}{2}\s^{g}{}_{(\a\b|}H^{(2)}{}_{g|\g)d}=~~~~
\s^{g}_{(\a\b|}\textbf{[}+2i
\g\nabla_{|\g)}(H^{(0)}{}_{def}H^{(0)}{}_{g}{}^{ef})~-~~
{2i\g}R^{(1)}{}_{|\g)[d|}{~}^{ef}H^{(0)}{}_{|g]ef}~~\nn\\
-~2i\g\Pi^{(1)}{}_{g}{}^{ef}[R^{(0)}{}_{|\g)def}~-
~2\nabla_{|\g)}H^{(0)}{}_{def}]\textbf{]}~~\nn\\
+~\frac{i\g}{24}\s^{pqref}{}_{(\a\b|}A^{(1)}{}_{pqr}
[R^{(0)}{}_{|\g)d}{}_{ef}~-~2\nabla_{|\g)}H^{(0)}{}_{def}]~\nn\\
-~~\frac{i}{2}\s_{d(\a|\la}T^{(2)}_{|\b\g)}{}^{\la}~
+~~\s^{g}{}_{(\a\b|}\frac{i}{2}T^{(2)}{}_{d|\g) g}~~ \eea

We write the expression this way for later convenience and
transparency. Now we consider the sigma five part. Although we do
not do it now, we note that the term with $R^{(0)}{}_{\g def}$
allows it to be written as a solvable term because of the identity,
(18). After a lengthy calculation we find the following (see \cite{6} for the detailed calculation ):

 \bea
\frac{+i\g}{24}\s^{pqref}{}_{(\a\b|}A^{(1)}{}_{pqr}[R^{(0)}{}_{|\g)def}~-
~2\nabla_{|\g)}H^{(0)}{}_{def}]
=~-\frac{\g}{4}\s^{g}_{(\a\b|}A^{(1)}{}_{gef}\s_{d|\g)\la}T^{(0)}{}^{ef}{}^{\la}~~\nn\\
+~4i\g^{2}\s_{e(\a|\eps}\s_{f(\b|\tau}\s_{d|\g)\la}T^{(0)}{}^{ef}{}^{\la}T^{(0)}{}_{kp}{}^{\eps}T^{(0)}{}^{kp}{}^{\tau}
=+~\frac{\g}{24}\s^{pqref}{}_{(\a\b|}A^{(1)}{}_{pqr}\s_{d|\g)\f}T_{ef}{}^{\f}
~~\eea

We now make the following identification:
 \bea
i\s_{d(\a|\la}T^{(2)}{}_{|\b\g)}{}^{\la}=+\frac{i\g}{12}
\s^{pqref}{}_{(\a\b|}A^{(1)}{}_{pqr}[R^{(0)}{}_{|\g)d}{}_{ef}-2\nabla_{|\g)}H^{(0)}{}_{def}]\nn\\
\eea or
 \bea
T^{(2)}{}_{\a\b}{}^{\la}~=~-~\frac{i\g}{12}\s^{pqref}{}_{\a\b}A^{(1)}{}_{pqr}T_{ef}{}^{\la}
\eea

This scenario will give for $H^{(2)}{}_{g \g d}$

\bea \frac{i}{2}\s^{g}{}_{(\a\b|}H^{(2)}{}_{g|\g)d}=~~~~
\s^{g}{}_{(\a\b|}\textbf{[}+2i
\g\nabla_{|\g)}(H^{(0)}{}_{def}H^{(0)}{}_{g}{}^{ef})~-~~
{2i\g}R^{(1)}{}_{|\g)[d|}{}^{ef}H^{(0)}{}_{|g]ef}~~\nn\\
-~2i\g [\Pi^{(1)}{}_{g}{}^{ef}][R^{(0)}{}_{|\g)def}~-
~2\nabla_{|\g)}H^{(0)}{}_{def}]\textbf{]}~~\nn\\
+~~\s^{g}{}_{(\a\b|}\frac{i}{2}T^{(2)}{}_{d|\g) g}~~ \eea

This is now in the solvable form and the result can be extracted
using the operator (77). It is written in terms of the torsion
$T^{(2)}{}_{d \g g}$, which is given later in (81).

\section{Closure Via the X tensor}

We now propose an Ansatz for the X tensor and show that, in
conjunction with the results (16) and (30), it succeeds in closing
the H sector by a different route. Furthermore we can add support to
this Ansatz by deriving the torsion $T^{(2)}{}_{\b\g}{}^{\la}$
through solving the dimension one-half torsion. Let us consider the
last two terms in (28), consisting of torsions.

Using the dimension one-half torsion (6), we can write the last two
terms of (28) as \bea
-~~\frac{i}{2}\s_{d(\a|\la}T^{(2)}_{|\b\g)}{}^{\la}~
+~~\s^{g}{}_{(\a\b|}\frac{i}{2}T^{(2)}{}_{d|\g) g}~~\nn\\
~=~\frac{1}{2}T^{(0)}{}_{(\a\b|}{}^{\la}T^{(2)}{}_{|\g)\la}{}^{d}~
-~\frac{1}{2}\nabla_{(\a|}T^{(2)}{}_{|\b\g)}{}^{d} \eea

Let the X tensor be given by the following:

\bea X_{pqrefd} =-\frac{i\g}{6}
H^{(0)}{}_{def}A^{(1)}{}_{pqr}+Y_{pqrefd} \eea

 Our procedure is successful with $Y_{pqrefd}=0$, hence we use
 \bea
T^{(2)}{}_{\a\b}{}^{d}=-~\frac{i\g}{6}\s^{pqref}{}_{\a\b}
H^{(0)}{}^{d}{}_{ef}A^{(1)}{}_{pqr}\eea

We obtain

 \bea -~\frac{i}{2}\s_{d(\a|\la}T^{(2)}_{|\b\g)}{}^{\la}~
+~\s^{g}{}_{(\a\b|}\frac{i}{2}T^{(2)}{}_{d|\g) g}~~\nn\\
~=~-\frac{i\g}{12}T^{(0)}{}_{(\a\b|}{}^{\la}[\s^{pqref}{}_{|\g)\la}
H^{(0)}{}_{def}A^{(1)}{}_{pqr}]~
+~\frac{i\g}{12}\s^{pqref}{}_{(\a\b|}
H^{(0)}{}_{def}[\nabla_{|\g)}A^{(1)}{}_{pqr}]\nn\\
+~\frac{i\g}{12}\s^{pqref}{}_{(\a\b|}
[\nabla_{|\g)}H^{(0)}{}_{def}]A^{(1)}{}_{pqr} \nn\\
\eea

 Now apply the result (16) to (36) to obtain

 \bea -~\frac{i}{2}\s_{d(\a|\la}T^{(2)}_{|\b\g)}{}^{\la}~
+~\s^{g}{}_{(\a\b|}\frac{i}{2}T^{(2)}{}_{d|\g) g}~~\nn\\
~=-~\frac{i\g}{2}\s^{g}{}_{(\a\b|}[\nabla_{|\g)}A^{(1)}{}_{gef}]H^{(0)}{}_{d}{}^{ef}\nn\\
+2i\g\s^{g}{}_{(\a\b|}R^{(1)}{}_{|\g)gef}H^{(0)}{}_{d}{}^{ef}\nn\\~~\eea

Hence also using (18) we find

\bea~\frac{i}{2}\s^{g}{}_{(\a\b|}H^{(2)}{}_{g|\g)d}~=~\s^{g}_{(\a\b|}\{+2i
\g\nabla_{|\g)}(H^{(0)}{}_{def}H^{(0)}{}_{g}{}^{ef})~-~
{2i\g}R^{(1)}{}_{|\g) [d|}{~}^{ef}H^{(0)}{}_{|g] ef}~~\nn\\
-~2i\g[\Pi^{(1)}{}_{g}{}^{ef}][R^{(0)}{}_{|\g)def}~-
~2\nabla_{|\g)}H^{(0)}{}_{def}]\}~~\nn\\
+\frac{\g}{12}\s^{g}{}_{(\a\b|}\s^{pqre}{}_{g}{}_{|\g)\f}A^{(1)}{}_{pqr}T^{(0)}{}_{d
e}^{}{\f}~-~\frac{i\g}{2}\s^{g}{}_{(\a\b|}[\nabla_{|\g)}A^{(1)}{}_{gef}]H^{(0)}{}_{d}{}^{ef}
+2i\g\s^{g}{}_{(\a\b|}R^{(1)}{}_{|\g)gef}H^{(0)}{}_{d}{}^{ef}\nn\\~~\eea

Thus we see that we can solve for $H^{(2)}{}_{g \g d}$ using this X
tensor in conjunction with our result for $T^{(2)}{}_{\b\g
}{}^{\la}$.

We note that we have also found the torsion
$T^{(2)}{}_{\b\g}{}^{g}$, by comparing (32) to (38). However for
clarity we will solve the dimension one-half torsion directly.
Looking at the dimension one-half torsion (6), and substituting (30)
and (35) into it, yields

\bea
-\frac{i\g}{6}T^{(0)}_{(\a\b|}{}^{\la}\s^{pqref}{}_{|\g)\la}A^{(1)}{}_{pqr}H^{(0)}{}_{def}
+~\frac{i\g}{12}\s^{pqref}{}_{(\a\b|}A^{(1)}{}_{pqr}
R^{(0)}{}_{\g)}{}_{def}\nn\\
~-~\frac{i\g}{6}\s^{pqref}{}_{(\a\b|}A^{(1)}{}_{pqr}\nabla_{|\g)}H^{(0)}{}_{def}
~+~\frac{i\g}{6}\s^{pqref}{}_{(\a\b|}A^{(1)}{}_{pqr}\nabla_{|\g)}H^{(0)}{}_{def}\nn\\
+~\frac{i\g}{6}\s^{pqref}{}_{(\a\b|}H^{(0)}{}_{def}[\nabla_{|\g)}A^{(1)}{}_{pqr}]\nn\\
-i\s^{g}{}_{(\a\b|}T^{(2)}{}_{|\g)}{}_{gd}=0 \eea

We now follow the same procedure as before and use (16) and (18). We
obtain convenient cancelations of non solvable terms to get

\bea +~i\s^{g}{}_{(\a\b|}T^{(2)}{}_{|\g) gd}~=
-i\g\s^{g}{}_{(\a\b|}[\nabla_{|\g)}A^{(1)}{}_{gef}]H^{(0)}{}_{d}{}^{ef}\nn\\
+4i\g\s^{g}{}_{(\a\b|}R^{(1)}{}_{|\g)gef}H^{(0)}{}_{d}{}^{ef}
~+~\frac{\g}{6}\s^{g}{}_{(\a\b|}\s^{pqre}{}_{g}{}_{|\g)\f}A^{(1)}{}_{pqr}T^{(0)}{}_{de}{}^{\f}
\eea

To find $T^{(2)}{}_{\g gd}$ the above must be symmetrized. The
 expression is listed in the conclusions, see eq. (81), and so is the spinor
 derivative $\nabla_{\g}A^{(1)}{}_{gef}$.

\section{Direct Method}

In the following we show agreement and consistency with the previous
results, by choosing to take the direct derivative of
$H^{(2)}{}_{\a\b d}$ (equation (20)), instead of eliminating non
linear terms using a torsion and curvature. This will involve the
derivative of $L^{(1)}{}_{abc}$. We begin with equation (21). We
require the following:
 \bea \nabla_{\g }\Pi^{(1)}{}_{g}{}^{ef}=
\nabla_{\g}[L^{(1)}{}_{g}{}^{ef}-\frac{1}{8}A^{(1)}{}_{g}{}^{ef}]\eea
We also have the important observation that, using
$\s^{g}{}_{(\a\b|}\s_{g |\g)\la}=0$, we can make the identification

\bea
 \nabla_{\a}L^{(1)}{}_{abc}=R^{(1)}{}_{\g abc}
\eea With this we can compare the two results. Taking the derivative
of (20) with the use of this result yields \bea
{\frac{1}{2}}\nabla_{(\a|}H^{(2)}{}_{|\b\g)d}{}^{Order\g^{2}}=
\s^{g}{}_{(\a\b|}[2i\g\nabla_{|\g)}(H^{(0)}{}_{def}H^{(0)}{}_{g}{}^{ef})^{Order(2)}~
+~4i\g\nabla_{|\g)}[H^{(0)}{}_{def}]\Pi^{(1)}{}_{g}{}^{ef}\nn\\
+4i\g
H^{(0)}{}_{def}R^{(1)}{}_{|\g)g}{}^{ef}-\frac{i\g}{2}\nabla_{|\g)}[A^{(1)}{}_{g}{}^{ef}]]
~-\frac{i\g}{24}\s^{pqref}{}_{(\a\b|}[\nabla_{|\g)}H^{(0)}{}_{def}]A^{(1)}{}_{pqr}\nn\\
~-\frac{i\g}{24}\s^{pqref}{}_{(\a\b|}H^{(0)}{}_{def}[\nabla_{|\g)}A^{(1)}{}_{pqr}]
~\nn\\
~~\eea

Substituting (43) as well as (42) into (21) yields exactly (32) with
the expression $\s^{g}{}_{(\a\b|}T^{(2)}{}_{|\g) gd}$ as found in
(40). We regard this as an important check for consistency.

\section{Torsion Equation for $T^{(2)}{}_{\a d}{}^{\d}$}

To second order we have the following torsion equation:

\bea T^{(0)}{}_{(\a\b|}{}^{\la}T^{(2)}{}_{|\g)\la}{}^{\d}~+
~T^{(2)}{}_{(\a\b|}{}^{\la}T^{(0)}{}_{|\g)\la}{}^{\d}
-i\s^{g}{}_{(\a\b|}T^{(2)}{}_{|\g)g}{}^{\d}~-~\nabla_{(\a|}T^{(2)}{}_{|\b\g)}{}^{\d}\nn\\
-~\frac{1}{4}R^{(2)}{}_{(\a\b|
de}\s^{de}{}_{|\g)}{}^{\d}~=~0\nn\\\eea

We must take care not to neglect second order contributions from the
derivative $\nabla_{\a}T^{(0)}{}_{\b\g}{}^{\d}$. We find

\bea -\nabla_{(\a|}T^{(2)}{}_{|\b\g)}{}^{\d}=
[2\d_{(\a|}{}^{\d}\d_{|\b)}{}^{\la}+\s^{g}{}_{(\a\b|}\s_{g}{}^{\d
\la}]\nabla_{|\g)}\chi_{\la}=\nn\\
-\frac{i}{2}\s^{g}{}_{(\a\b|}\s^{mn}{}_{|\g)}{}^{\d}[L^{(2)}{}_{gmn}+\frac{1}{4}A^{(2)}{}_{gmn}]
\eea  Once again using (12), (16) and (31), reduces (44) to
\newpage
\bea
-\frac{i\g}{2}\s^{g}{}_{(\a\b|}[\nabla_{|\g)}A^{(1)}{}_{g}{}^{ef}]T^{(0)}{}_{ef}{}^{\d}
+2i\g\s^{g}{}_{(\a\b|}R^{(1)}{}_{|\g) g}{}^{ef}T^{(0)}{}_{ef}{}^{\d}\nn\\
~-~\frac{i\g}{12}\s^{pqref}{}_{(\a\b|}A^{(1)}{}_{pqr}T^{(0)}{}_{ef}{}^{\la}T^{(0)}{}_{|\g)\la}{}^{\d}
-i\s^{g}{}_{(\a\b|}T^{(2)}{}_{|\g)g}{}^{\d}\nn\\
+~\frac{i\g}{12}\s^{pqref}{}_{(\a\b|}A^{(1)}{}_{pqr}[-\frac{1}{4}\s^{mn}{}_{|\g)}{}^{\d}R_{efmn}\nn\\
~+~T^{(0)}{}_{ef}{}^{\la}T^{(0)}{}_{|\g)\la}{}^{\d}]-~\frac{1}{4}R^{(2)}{}_{(\a\b|de}\s^{de}{}_{|\g)}{}^{\d}~\nn\\
-\frac{i}{2}\s^{g}{}_{(\a\b|}\s^{mn}{}_{|\g)}{}^{\d}[L^{(2)}{}_{gmn}+\frac{1}{4}A^{(2)}{}_{gmn}]~=~0~
~\eea We note that the spinor derivative of $T^{(0)}{}_{ef}{}^{\d} $
results in a convenient cancelation of otherwise unsolvable terms,
and so finally we obtain
 \bea
+i\s^{g}{}_{(\a\b|}T^{(2)}{}_{|\g)g}{}^{\d}\nn\\
~+\frac{i\g}{2}\s^{g}{}_{(\a\b|}[\nabla_{|\g)}A^{(1)}{}_{g}{}^{ef}]T^{(0)}{}_{ef}{}^{\d}
-2i\g \s^{g}{}_{(\a\b|}R^{(1)}{}_{|\g) g}{}^{ef}T^{(0)}{}_{ef}{}^{\d}\nn\\
+~\frac{1}{4}R^{(2)}{}_{(\a\b| de}\s^{de}{}_{|\g)}{}^{\d}
+~\frac{i\g}{12}\s^{pqref}{}_{(\a\b|}A^{(1)}{}_{pqr}[+\frac{1}{4}\s^{mn}{}_{|\g)}{}^{\d}R_{efmn}]\nn\\
+\frac{i}{2}\s^{g}{}_{(\a\b|}\s^{mn}{}_{|\g)}{}^{\d}[L^{(2)}{}_{gmn}+\frac{1}{4}A^{(2)}{}_{gmn}]
=0\eea

From the above we extract the candidates

\bea T^{(2)}{}_{\g g}{}^{\d} =~- \frac{\g}{2}
[\nabla_{\g}A^{(1)}{}_{gef}]T^{(0)}{}^{ef}{}^{\d}+2\g
R^{(1)}{}_{\g gef}T^{(0)}{}^{ef}{}^{\d}\nn\\
~~\eea and \bea R^{(2)}{}_{\a\b de}~=~
-~\frac{i\g}{12}\s^{pqrab}{}_{\a\b}A^{(1)}{}_{pqr}R_{deab}-
2i\s^{g}{}_{\a\b}[L^{(2)}{}_{gde}+\frac{1}{4}A^{(2)}{}_{gde}] \eea

\section{Curvature Equation for $R^{(2)}{}_{\la gde}$}

 We need to solve the curvature that
will give $R^{(2)}{}_{\la gde}$. We have the curvature Bianchi
identity which we need to solve at second order as follows:

\bea
T_{(\a\b|}{}^{\la}R_{|\g)\la}{}_{de}~-~T_{(\a\b|}{}^{g}R_{|\g)}{}_{gde}~
-~\nabla_{(\a|}R_{\b\g)}{}_{de}~=~0\eea

Here we must consider second order contributions from the spinor
derivative $\nabla_{(\a|}R_{|\b\g)}{}_{de}$ . We write the full
curvature to second order for clarity

\bea
R_{\b\g}{}_{de}=~-2i\s^{g}{}_{\a\b}\Pi'_{gde}+~\frac{i}{24}\s^{pqr}{}_{de}{}_{\a\b}A^{(1)}{}_{pqr}
-~\frac{i\g}{12}\s^{pqrab}{}_{\a\b}A^{(1)}{}_{pqr}R_{deab} \eea

Where $\Pi'$ is the modified $\Pi$, but is any case is of the
solvable form.

\bea
\Pi'=[L^{(0)}{}_{gde}+L^{(1)}{}_{gde}+L^{(2)}{}_{gde}-\frac{1}{4}A^{(1)}{}_{gde}+\frac{1}{4}A^{(2)}{}_{gde}]\eea

With hindsight and in order to eliminate an apparently intractable
term we begin by making the following observations. In we found
$T^{(2)}{}_{\a\b}{}^{\la}$, hence we also encountered the quantity

 \bea
\s_{d(\a|\la}T^{(2)}{}_{\a\b}{}^{\la}~=~-\frac{i\g}{12}\s_{d(\a|\la}\s^{pqref}{}_{|\b\g)}A^{(1)}{}_{pqr}T^{(0)}_{ef}{}^{\la}
\eea Using the torsion, equation (6) we can write

\bea \s_{d(\a|\la}T^{(2)}_{|\b\g)}{}^{\la}~
=-~~\s^{g}{}_{(\a\b|}T^{(2)}{}_{d|\g) g}~~\nn\\
+~iT^{(0)}{}_{(\a\b|}{}^{\la}T^{(2)}{}_{|\g)\la}{}^{d}~
-i\nabla_{(\a|}T^{(2)}{}_{|\b\g)}{}^{d} \eea

\noindent Using the second order torsion results which we found then
gives

 \bea +\s_{d(\a|\la}T^{(2)}_{|\b\g)}{}^{\la}~=
+~\s^{g}{}_{(\a\b|}T^{(2)}{}_{d|\g) g}~~\nn\\
~~+\frac{\g}{6}T^{(0)}{}_{(\a\b|}{}^{\la}[\s^{pqref}{}_{|\g)\la}
H^{(0)}{}_{def}A^{(1)}{}_{pqr}]~ -~\frac{\g}{6}\s^{pqref}{}_{(\a\b|}
H^{(0)}{}_{def}[\nabla_{|\g)}A^{(1)}{}_{pqr}]\nn\\
-~\frac{\g}{6}\s^{pqref}{}_{(\a\b|}
[\nabla_{|\g)}H^{(0)}{}_{def}]A^{(1)}{}_{pqr} \nn\\
\eea

Now applying our key equation, (16) to (55) gives

 \bea +\s_{d(\a|\la}T^{(2)}_{|\b\g)}{}^{\la}~=
+~\s^{g}{}_{(\a\b|}T^{(2)}{}_{d|\g) g}~~\nn\\
~+~\g\s^{g}{}_{(\a\b|}[\nabla_{|\g)}A^{(1)}{}_{gef}]H^{(0)}{}_{d}{}^{ef}
-4\g\s^{g}{}_{(\a\b|}R^{(1)}{}_{|\g)gef}H^{(0)}{}_{d}{}^{ef}\nn\\~~\eea

We now substitute in out result for
$\s^{g}{}_{(\a\b|}T^{(2)}{}_{d|\g) g}$, (40) into (56) to obtain
cancelations and the simple result

\bea \s_{d(\a|\la}T^{(2)}_{|\b\g)}{}^{\la}~=
-~\frac{i\g}{6}\s^{g}{}_{(\a\b|}\s^{pqre}{}_{g}{}_{|\g)\f}A^{(1)}{}_{pqr}T^{(0)}{}_{de}{}^{\f}\nn\\
=-\frac{i\g}{12}\s_{d(\a|\la}\s^{pqref}{}_{|\b\g)}A^{(1)}{}_{pqr}T^{(0)}_{ef}{}^{\la}
\nn\\~~\eea

From which we extract the following result \cite{6}

\bea
\s^{pqref}{}_{(a\b|}\s_{d|\g)|\f}A^{(1)}{}_{pqr}T^{(0)}_{ef}{}^{\f}
=2\s^{g}{}_{(\a\b|}\s^{pqre}{}_{g}{}_{|\g)\f}A^{(1)}{}_{pqr}T^{(0)}{}_{de}{}^{\f}\nn\\
\nn\\~~\eea

From this we deduce a hitherto unknown identity, albeit indirectly. Using our second order torsion and curvature results,
(31), (35), and (51), the full curvature at second order becomes

 \bea ~T^{(0)}_{(\a\b|}{}^{\la}[ -~\frac{i\g}{12}\s^{pqrab}{}_{|\g)\la}A^{(1)}{}_{pqr}R^{(0)}{}_{abde}]~
 +~[-~\frac{i\g}{12}\s^{pqrab}{}_{(\a\b|}A^{(1)}{}_{pqr}T_{ab}{}^{\la}R^{(0)}{}_{|\g)\la}{}_{de}\nn\\
~-~i\s^{g}{}_{(\a\b|}R^{(2)}{}_{|\g)}{}_{gde}~+~\frac{i\g}{6}\s^{pqrab}{}_{(\a\b|}
H^{(0)}{}^{g}{}_{ab}A^{(1)}{}_{pqr}]R^{(0)}{}_{|\g)}{}_{gde}\nn\\
+~\frac{i\g}{12}\s^{pqrab}{}_{(\a\b|}[\nabla_{|\g)|}A^{(1)}{}_{pqr}]R^{(0)}{}_{abde}
+~\frac{i\g}{12}\s^{pqrab}{}_{(\a\b|}A^{(1)}{}_{pqr}[\nabla_{|\g)}R^{(0)}{}_{abde}]~\nn\\
-~\frac{i}{24}\s^{pqr}{}_{de}{}_{(\a\b|}[\nabla_{|\g)}A^{(1)}{}_{pqr}]
+~2i\g\s^{g}{}_{(\a\b|}[\nabla_{|\g)}\Pi'{}_{gde}] =~0 \eea

Using (16) again gives two more solvable terms

\bea
~+\frac{i\g}{12}T^{(0)}{}_{(\a\b|}{}^{\la}\s^{pqr}{}_{ab|\g)\la}A^{(1)}{}_{pqr}R^{(0)}{}^{ab}{}_{de}-
\frac{i\g}{12}\s^{pqrab}{}_{(\a\b|}R^{(0)}{}_{abde}\nabla_{|\g)}A^{(1)}{}_{pqr}\nn\\
=+\frac{i\g}{2}\s^{g}{}_{(\a\b|}[\nabla_{|\g)}A^{(1)}{}_{gab}]R^{(0)}{}^{ab}{}_{de}-
2i\g\s^{g}{}_{(\a\b|}R^{(1)}{}_{|\g)
g}{}^{ab}R^{(0)}{}_{abde}\nn\\\eea

This reduces (59) to
 \bea~~+\frac{i\g}{2}\s^{g}{}_{(\a\b|}\nabla_{|\g)}A^{(1)}{}_{gab}R^{(0)}{}^{ab}{}_{de}-
2i\g\s^{g}{}_{(\a\b|}R^{(1)}{}_{|\g) gab}R^{(0)}{}_{abde}\nn\\
-~\frac{i\g}{12}\s^{pqrab}{}_{(\a\b|}A^{(1)}{}_{pqr}T_{ab}{}^{\la}R^{(0)}{}_{|\g)\la}{}_{de}\nn\\
~-~i\s^{g}{}_{(\a\b|}R^{(2)}{}_{|\g)}{}_{gde}~+~\frac{i\g}{6}\s^{pqrab}{}_{(\a\b|}
H^{(0)}{}^{g}{}_{ab}A^{(1)}{}_{pqr}R^{(0)}{}_{|\g)}{}_{gde}\nn\\
+~\frac{i\g}{12}\s^{pqrab}{}_{(\a\b|}A^{(1)}{}_{pqr}[\nabla_{|\g)}R^{(0)}{}_{abde}]~\nn\\
-~\frac{i}{24}\s^{pqr}{}_{de}{}_{(\a\b|}[\nabla_{|\g)}A^{(1)}{}_{pqr}]
+~2i\g\s^{g}{}_{(\a\b|}[\nabla_{|\g)}\Pi'{}_{gde}]=~0\nn\\
\eea

We now list the sigma five terms separately.

\bea +\frac{i\g}{12}\s^{pqrab}{}_{(\a\b|}A^{(1)}{}_{pqr}[
-T^{(0)}{}_{ab}{}^{\la}R^{(0)}{}_{\la|\g)}{}_{de}~
+~2H^{(0)}{}_{ab}{}^{g}R^{(0)}{}_{|\g)}{}_{gde}~+~\nabla_{|\g)}R^{(0)}{}_{abde}]
\eea

\bea =+\frac{i\g}{12}\s^{pqrab}{}_{(\a\b|}A^{(1)}{}_{pqr}[
-T^{(0)}{}_{ab}{}^{\la}R^{(0)}{}_{\la|\g)}{}_{de}~-~T^{(0)}{}_{ab}{}^{g}R^{(0)}{}_{\g}{}_{gde}~
+~\nabla_{|\g)}R^{(0)}{}_{abde}] \eea

 We have the Bianchi Identity

\bea \nabla_{\a}R_{abde}~-~T_{\a [a|}{}^{X}R_{X}{}_{|b]d e}~
-~T_{ab}{}^{X}R_{X}{}_{\a de}~+~\nabla_{[a|}R_{|b]\a d e}=0 \eea

The second term on the LHS of  (64) is zero at zeroth order. Hence
we have as follows:

\bea
+\frac{i\g}{12}\s^{pqrab}{}_{(\a\b|}A^{(1)}{}_{pqr}[\nabla_{|\g)}R^{(0)}{}_{abde}]=\nn\\
+\frac{i\g}{12}\s^{pqrab}{}_{(\a\b|}A^{(1)}{}_{pqr}[+T^{(0)}{}_{ab}{}^{\la}R^{(0)}{}_{\la|\g)}{}_{de}~
-~T^{(0)}{}_{ab}{}^{g}R^{(0)}{}_{|\g)}{}_{gde}~-~2\nabla_{a}R^{(0)}{}_{b|\g)de}]
\eea

 Substituting (65) into (61) gives

\bea~+\frac{i\g}{2}\s^{g}{}_{(\a\b|}[\nabla_{|\g)}A^{(1)}{}_{gab}]R^{(0)}{}^{ab}{}_{de}-
2i\g\s^{g}{}_{(\a\b|}R^{(1)}{}_{|\g) gab}R^{(0)}{}_{abde}\nn\\
~-~i\s^{g}{}_{(\a\b|}R^{(2)}{}_{|\g)}{}_{gde}~
+~\frac{i\g}{6}\s^{pqrab}{}_{(\a\b|}A^{(1)}{}_{pqr}[\nabla_{a}R^{(0)}{}_{|\g)bde}+
2H^{(0)}_{ab}{}^{g}R^{(0)}_{|\g)gde}]
\nn\\-~\frac{i}{24}\s^{pqr}{}_{de}{}_{(\a\b|}[\nabla_{|\g)}A^{(1)}{}_{pqr}]
+~2i\g\s^{g}{}_{(\a\b|}[\nabla_{|\g)}\Pi'{}_{gde}]=0\eea

Now consider the sigma five terms in (66). Using our new result
result (58) allows for solving these terms, and we obtain \bea
+~\frac{\g}{6}\s^{pqrab}{}_{(\a\b|}A^{(1)}{}_{pqr}[\nabla_{a}R^{(0)}{}_{|\g)bde}+
2H^{(0)}_{ab}{}^{g}R^{(0)}_{|\g)gde}]\nn\\
=~\frac{\g}{6}\s^{pqrab}{}_{(\a\b|}A^{(1)}{}_{pqr}[\s_{[d|g)\f}\{\nabla_{a}T^{(0)}{}_{b|e]}{}^{f}
+2H^{(0)}_{ab}{}^{c}T^{(0)}{}_{c|e]}{}^{\f}\}] \nn\\
= ~\frac{\g}{3}\s^{g}{}_{(\a\b|}\s^{pqra}{}_{g}{}_{|\g)\f}
A^{(1)}{}_{pqr}[\{\nabla_{[d|}T^{(0)}{}_{a|e]}{}^{f}
+2H^{(0)}_{[d|a}{}^{c}T^{(0)}{}_{c|e]}{}^{\f}\}] \eea

Hence we obtain

\bea~ ~i\s^{g}{}_{(\a\b|}R^{(2)}{}_{|\g)}{}_{gde}~=
+\frac{i\g}{2}\s^{g}{}_{(\a\b|}[\nabla_{|\g)}A^{(1)}{}_{gab}]R^{(0)}{}^{ab}{}_{de}-
2i\g\s^{g}{}_{(\a\b|}R^{(1)}{}_{|\g) gab}R^{(0)}{}_{abde}\nn\\
+~\frac{\g}{3}\s^{g}{}_{(\a\b|}\s^{pqra}{}_{g}{}_{|\g)\f}
A^{(1)}{}_{pqr}[\{\nabla_{[d|}T^{(0)}{}_{a|e]}{}^{\f}
+2H^{(0)}_{[d|a}{}^{c}T^{(0)}{}_{|c|e]}{}^{\f}\}]
\nn\\-~\frac{i}{24}\s^{pqr}{}_{de}{}_{(\a\b|}[\nabla_{|\g)}A^{(1)}{}_{pqr}]
+~2i\g\s^{g}{}_{(\a\b|}[\nabla_{|\g)}\Pi'{}_{gde}]=0\eea

We now look at the remaining unsolved term
$-~\frac{i}{24}\s^{pqr}{}_{de}{}_{(\a\b|}[\nabla_{|\g)}A^{(1)}{}_{pqr}]$.
This term cannot be manipulated into a solvable term because of the
placement of the free indices. Using the results found in \cite{5}
we have

 \bea-~\frac{i}{24}\s^{pqr}{}_{de}{}_{(\a\b|}[\nabla_{|\g)}A^{(1)}{}_{pqr}]=\nn\\
+\frac{\g}{(12)(24)}\s^{pqr}{}_{de(\b\g|}\s_{pqr
\eps\tau}T^{(0)}{}^{\eps}{}_{kl}\s^{mns}{}^{\f}{}^{\tau}[H^{(0)}{}^{kl}{}_{g}\s^{g}{}_{|\a)}{}_{\f}A^{(1)}{}_{mns}\nn\\
-\s_{k}{}_{|\a)}{}_{\f}(\nabla_{l}A^{(1)}{}_{mns})] \eea

Using the sigma matrix identities as given in (3) it can be shown
that these two terms cannot be written in the solvable form, that is
with the same structure as
$\s^{g}{}_{(\a\b|}R^{(2)}{}_{|\g)}{}_{gde}$. Hence we look at the
origin of these terms. For the derivative of $T_{kl}{}^{\tau}$ we
have the following Bianchi identity.

\bea
\nabla_{\g}T_{kl}{}^{\tau}=T_{\g[k|}{}^{\la}T_{\la|l]}{}^{\tau}+
T_{\g[k}{}^{g}T_{g|l]}{}^{\tau}+T_{kl}{}^{\la}T_{\la
\g}{}^{\tau}+T_{kl}{}^{g}T_{g\g}{}^{\tau}-\nabla_{[k|}T_{|l]\g}{}^{\tau}
-R_{kl\g}{}^{\tau}\eea

At first order this reduces to

\bea \nabla_{\g}T_{kl}{}^{\tau}{}^{Order (1)}=
T^{(1)}{}_{\g[k|}{}^{\la}T^{(0)}{}_{\la|l]}{}^{\tau}-\nabla_{[k|}T_{|l]\g}{}^{\tau}{}^{Order(1)}
-R^{(1)}{}_{kl\g}{}^{\tau}\eea

In references \cite{1} and \cite{5} it appears that $R^{(1)}{}_{kl
\g}{}^{\tau}$ was set to zero. With the form of the curvature
$R_{\a\b de}$ and this choice of super current supertensor $A_{abc}$
we will always be led to the term
$\frac{i}{24}\s^{pqr}{}_{de}{}_{(\a\b|}\nabla_{|\g)}A^{(1)}{}_{pqr}{}^{(order2)}$
because of the the spinor derivative in the Bianchi identity (32) as
given in (45). This term is not reducible as we require so it must
be incorporated into this curvature. Hence we must identify the
following curvature at first order:

\bea R^{(1)}{}_{kl\g}{}^{\tau}=\frac{1}{48}
[2H^{(0)}{}_{kl}{}_{g}\s^{g}{}_{\g
\la}\s^{pqr\la\tau}A^{(1)}{}_{pqr}
-\s_{[k|\g\la}\s^{pqr\la\tau}{}(\nabla_{|l]}A^{(1)}{}_{pqr})]\eea

The second order form of this curvature is already solved in the
Bianchi identity (71). All the quantities in this Bianchi identity
are known. Hence it can be written in full in a later review. It is
the role of this paper simply to arrive at the second order
solution, and to overcome obstacles to obtaining this solution.
\section{The Super-current}
The starting point in references \cite{1} and \cite{2}were the
conventional constraints as listed in \cite{1}. Among these
constraints we have

\bea T_{\a b}{}^{\d}=\frac{1}{48}\s_{b\a \la}\s^{pqr
\la\d}A_{pqr}\eea
The choice of

\bea A_{pqr}=-i\g \s_{pqr \eps \tau}T_{kp}{}^{\eps}T^{kp \tau}\eea
 was made for on
shell conditions, [1]. This conventional constraint can be imposed
to all orders. We have found $T_{\a b}{}^{\d}$ and it is given in in
equation (48). Hence we can find $A^{(2)}{}_{pqr}$, by solving the
above \cite{6}. No modification to this super-current was required
to close the identities other than this Hence we use a suitable
inverting operator along with our results (42) and (48) to obtain

\bea A^{(2)}{}_{gef}=-\frac{1}{20}\s_{gef \g\la}\s^{b \la
\f}T^{(2)}{}_{\f b}{}^{\g} \eea

\bea =\frac{\g}{20}\s_{gef \g\la}\s^{b \la
\f}[\nabla_{\f}(\frac{1}{4}A^{(1)}{}_{bmn}
-2L^{(1)}{}_{bmn})]T^{(0)mn \g}\eea

\section{Conclusions}

We have analyzed the second order non minimal case of string
corrected supergravity. We found a procedure for solving the Bianchi
identities to this order. This involved the equations (16), (31),
(35), (42), (58), and (72), which we used in conjunction with
several other key observations. We found a mechanism which allows
for closure of the H sector Bianchi identities and also related
torsions and curvatures.

 We have seen how the X tensor is necessary for achieving closure
 of these identities and we have proposed a candidate for this tensor which succeeds.
Adding the second part, $Y_{pqrdef}$ ,to the X tensor appears to
result in failure to close in the H sector. With $H^{(2)}{}_{\a\b
\d}$ set to zero we obtained $H^{(2)}{}_{\a\b d}$, given in equation
(20).

$H^{(2)}{}_{\a b g}$ must be extracted from (32). We use the
following operator, $\hat{O}$,  to obtain the symmetrized
$H^{(2)}{}_{\a b g}$:

\bea \hat{O}=[\frac{1}{2}\d_{[a}{}^{d}\d_{b]}{}^{g}\d_{\a}{}^{\b}~
-~\frac{1}{12}\eta^{dg}\s_{ab\a}{}^{\b}~+\frac{1}{24}\d_{[a}{}^{(d}\s_{b]}{}^{g)}_{\a}{}^{\b}]
\eea  We find after a long calculation we obtain the result

\bea H^{(2)}_{\a
ab}=~2\g[\nabla_{\a}(H^{(0)}_{[a|ef}H^{(0)}{}_{|b]}{}^{ef}~
-~\s_{ab\a}{}^{\f}\nabla_{\f}(H^{(0)}{}_{gef}H^{gef})]\nn\\
~+2i\g\s_{[a|\a\f}T_{ef}{}^{\f}\Pi^{(1)}{}_{|b]}{}^{ef}
-~2i\g\s_{ab\a}{}^{\la}\s_{g\la\f}T_{ef}{}^{\f}\Pi^{(1)gef}\nn\\
-~\frac{\g}{6}\s^{g}{}_{[a|\a}{}^{\f}\s_{|b]\la\f}T_{ef}{}^{\la}\Pi^{(1)}{}_{g}{}^{ef}
-~\frac{\g}{6}\s^{g}{}_{[a|\a}{}^{\f}\s_{g\la\f}T_{ef}{}^{\la}\Pi^{(1)}{}_{|b]}{}^{ef}\nn\\
~-~4\g R^{(1)}{}_{\a[a|}{}^{ef}H^{(0)}{}_{|b]ef}+T^{(2)}{}_{\a
ab}\eea

\noindent where $T^{(2)}{}_{\a ab}$ is given in equation (81). In
the case of $H^{(2)}{}_{a b c}$, the Bianchi identity has already
given us the result. From the term $T_{\a\b}{}^{E}H_{Ecd}$ in
equation (5) we isolate an expression of the form
\bea
T^{(0)}{}_{\a\b}{}^{g}H^{(2)}{}_{gcd}~=~i\s_{\a\b}{}^{g}H^{(2)}{}_{gcd}~=~M_{\a\b cd}\nn\\
\eea The right hand side contains now known torsions and curvatures.
However they need only be substituted into (79) generating a long
expression. We then use the fact that \bea
\s_{a}{}_{\a\b}\s^{b}{}^{\a\b}~=~-~16\d^{b}{}_{a}\eea and solve for
$H^{(2)}{}_{gcd}$.

This is left for the next stage of work. We obtained the full set or
torsions and and curvatures, (31),(35), (48) and( 49).

Extracting the symmetrized torsion $T^{(2)}{}_{\g gd}$ from (40)
gives

 \bea T^{(2)}{}_{\g ab}=
-\frac{\g}{2}[\nabla_{\g}A^{(1)}{}_{[a|ef}]H^{(0)}{}_{|b]}{}^{ef}+
2\g R^{(1)}{}_{\g [a|ef}H^{(0)}{}_{|b]}{}^{ef}
-\frac{i\g}{12}\s^{pqrg}{}_{[a|\g\la}T^{(0)}{}_{|b]g}{}^{\la}A^{(1)}{}_{pqr}\nn\\
+\s_{ab~\g}{}^{\f}[+\frac{\g}{12}
(\nabla_{\f}A^{(1)}{}_{gef})H^{(0)}{}^{gef}+\frac{\g}{3}R^{(1)}{}_{\f~gef}H^{(0)}{}^{gef}
-\frac{i\g}{72}\s^{pqreg}{}_{\f\la}A^{(1)}{}_{pqr}T^{(0)}{}_{eg}{}^{\la}]\nn\\
+\s_{[a|}{}^{g}{}_{\g}{}^{\f}[-\frac{\g}{2}(\nabla_{\f}A^{(1)}{}_{|b]ef})H^{(0)}{}_{g}{}^{ef}
-\frac{\g}{2}(\nabla_{\f}A^{(1)}{}_{gef})H^{(0)}{}_{|b]}{}^{ef}\nn\\
-\frac{\g}{6}R^{(1)}{}_{\f~|b]ef}H^{(0)}{}_{g}{}^{ef}-\frac{\g}{6}R^{(1)}{}_{\f~gef}H^{(0)}{}_{|b]}{}^{ef}\nn\\
+\frac{i\g}{144}A^{(1)}{}_{pqr}[\s^{pqre}{}_{|b]}{}_{\f\la}T^{(0)}{}_{eg}{}^{\la}
+\s^{pqre}{}_{g~\f\la}T^{(0)}{}_{e|b]}{}^{\la}]]\nn\\
\eea

where \bea
\nabla_{\g}A^{(1)}{}_{gef}=i\g\s_{gef~\eps\tau}T^{(0)}{}_{kp}{}^{\eps}[2T^{(0)}{}^{kp}{}^{
\la}T^{(0)}{}_{\g\la}{}^{\tau}-\frac{1}{2}\s^{mn}{}_{\g}{}^{\tau}R^{(0)}{}^{kp}{}_{mn}]\nn\\
\eea

We also find the adjusted curvature $R{}_{kl\g}{}^{\tau}$. For  $
R^{(2)}{}_{\a\b de}$ we have reduced it to solvable form. After
imposing condition (72) we have
\bea~
~i\s^{g}{}_{(\a\b|}R^{(2)}{}_{|\g)}{}_{gde}~=
\s^{g}{}_{(\a\b|}[\frac{i\g}{2}[\nabla_{|\g)}A^{(1)}{}_{gab}]R^{(0)}{}^{ab}{}_{de}-
2i\g R^{(1)}{}_{|\g)gab}R^{(0)}{}_{abde}\nn\\
+~\frac{\g}{3}\s^{pqra}{}_{g}{}_{|\g)\f}
A^{(1)}{}_{pqr}\{\nabla_{[d|}T^{(0)}{}_{a|e]}{}^{\f}
+2H^{(0)}_{[d|a}{}^{c}T^{(0)}{}_{|c|e]}{}^{\f}\}+~2i\g\nabla_{|\g)}\Pi'{}_{gde}]=0\eea

\noindent $ R^{(2)}{}_{\a\b de}$ can be extracted from the above
result. Finally we have found  the supercurrent $A^{(2)}{}_{abc}$ as
given in equation (76).

\subsection*{Acknowledgements}
This work was partially supported by the Marie Curie Research
Training Network under contract MRTN-CT-2004-005104 Forces Universe.
We would like to thank S. J. Gates, Jr. for his constant critical interest
in this work.

\end{document}